\documentclass[aps,twocolumn,superscriptaddress,amsmath]{revtex4}

\usepackage{epsf,latexsym}
\usepackage{amssymb,amsmath}
\usepackage{times}
\usepackage[final]{pdfpages}

\newcommand{\bra}[1]{\langle #1 |}
\newcommand{\ket}[1]{| #1 \rangle}
\newcommand{\braket}[2]{\langle #1 | #2 \rangle}

\newcommand\V{{\cal V}}

\newcommand\p{{\sf p}}
\newcommand\w{{\sf w}}

\newcommand\tr{{\mbox{Tr\,}}}

\def\half{\tfrac{1}{2}}

\newcommand{\ignore}[1]{}

\newcommand{\be}{\begin{equation}}
\newcommand{\ee}{\end{equation}}
\newcommand{\ba}{\begin{eqnarray}}
\newcommand{\ea}{\end{eqnarray}}

\def\CC{{\rm\kern.24em \vrule width.04em height1.46ex depth-.07ex
    \kern-.30em C}}
\def\P{{\rm I\kern-.25em P}}

\def\RR{{\rm
         \vrule width.04em height1.58ex depth-.0ex
         \kern-.04em R}}

\def\bbbc{{\mathchoice {\setbox0=\hbox{$\displaystyle\rm C$}\hbox{\hbox
to0pt{\kern0.4\wd0\vrule height0.9\ht0\hss}\box0}}
{\setbox0=\hbox{$\textstyle\rm C$}\hbox{\hbox
to0pt{\kern0.4\wd0\vrule height0.9\ht0\hss}\box0}}
{\setbox0=\hbox{$\scriptstyle\rm C$}\hbox{\hbox
to0pt{\kern0.4\wd0\vrule height0.9\ht0\hss}\box0}}
{\setbox0=\hbox{$\scriptscriptstyle\rm C$}\hbox{\hbox
to0pt{\kern0.4\wd0\vrule height0.9\ht0\hss}\box0}}}}

\def\bbbz{{\mathchoice {\hbox{$\sf\textstyle Z\kern-0.4em Z$}}
{\hbox{$\sf\textstyle Z\kern-0.4em Z$}}
{\hbox{$\sf\scriptstyle Z\kern-0.3em Z$}}
{\hbox{$\sf\scriptscriptstyle Z\kern-0.2em Z$}}}}

\newcommand{\putfig}[2]{$$\leavevmode\hbox{\epsfxsize=#2 cm
   \epsffile{#1.eps}}$$}

\begin{document}

\title{Proposal for a quantum delayed-choice experiment}

\author{Radu Ionicioiu}
\affiliation{Centre for Quantum Science and Technology, Macquarie University, Sydney NSW 2109, Australia}
\affiliation{Institute for Quantum Computing, University of Waterloo, Waterloo, Ontario N2L 3G1, Canada}
\affiliation{Department of Applied Mathematics, University of Waterloo, Waterloo, Ontario N2L 3G1, Canada}

\author{Daniel R. Terno}
\affiliation{Department of Physics \& Astronomy, Macquarie University, Sydney NSW 2109, Australia}
\affiliation{Centre for Quantum Technologies, National University of Singapore, Singapore 117543}

\begin{abstract}
{\em Gedanken} experiments are important conceptual tools in the quest to reconcile our classical intuition with quantum mechanics and nowadays are routinely performed in the laboratory. An important open question is the quantum behaviour of the controlling devices in such experiments. We propose a framework to analyse quantum-controlled experiments and illustrate the implications by discussing a quantum version of Wheeler's delayed-choice experiment. The introduction of a quantum-controlled device (i.e., quantum beamsplitter) has several consequences. First, it implies that we can measure complementary phenomena with a single experimental setup, thus pointing to a redefinition of complementarity principle. Second, a quantum control allows us to prove there are no consistent hidden-variable theories in which ``particle'' and ``wave'' are realistic properties. Finally, it shows that a photon can have a morphing behaviour between ``particle" and ``wave"; this further supports the conclusion that ``particle" and ``wave" are not realistic properties but merely reflect how we 'look' at the photon. The framework developed here can be extended to other experiments, particularly to Bell-inequality tests.
\end{abstract}


\maketitle

\begin{figure}[floatfix]
\putfig{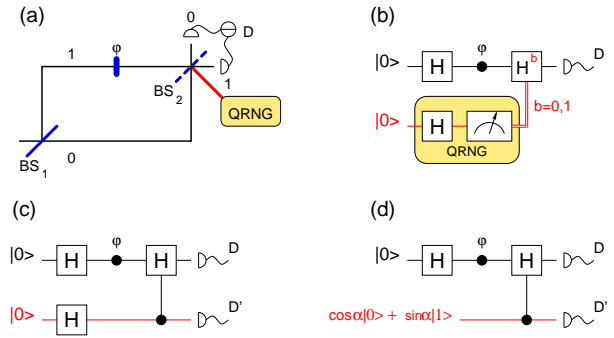}{8}
\caption{(a) In the classical delayed-choice experiment the second beam-splitter is inserted or removed randomly after the photon is already inside the interferometer. (b) The equivalent quantum network. An ancilla (red line), initially prepared in the state $\ket{+}= \frac{1}{\sqrt{2}}(\ket{0}+ \ket{1})$ then measured, acts as a quantum random number generator (QRNG). (c) Delayed-choice with a quantum beamsplitter. The classical control (red double line) after the measurement of the ancilla in (b) is equivalent to a quantum control before the measurement; the second beamsplitter BS$_2$ is now in superposition of present and absent, equivalent to a controlled-Hadamard $C(H)$ gate. (d) We bias the quantum random number generator (QRNG) by preparing the ancilla in an arbitrary state $\cos\alpha\ket{0}+ \sin\alpha\ket{1}$.}
\label{delayed_ch}
\end{figure}

Wave-particle duality, a quintessential property of quantum systems, defies our classical intuition. In the context of the double-slit experiment, duality played a central role in the famous Bohr--Einstein debate and prompted Bohr to formulate the complementarity principle \cite{bohr_wz}: ``the study of complementary phenomena demands mutually exclusive experimental arrangements". Classical concepts like ``particle" or ``wave" (as in ``wave-particle duality") do not translate perfectly into the quantum language. For example, although we observe interference (a definite wave-like behaviour), the pattern is produced click-by-click, in a discrete, particle-like manner \cite{gra}. Notwithstanding this ambiguity, and with this proviso, we adopt as operational definition of ``wave/particle" to stand for ``ability/inability to produce interference" \cite{green}.

A good illustration of wave-particle complementarity is given by a Mach-Zehnder interferometer (MZI), Figure \ref{delayed_ch}. A photon is first split by beamsplitter BS$_1$, travels inside an interferometer with a tunable phase shifter $\varphi$, and is finally recombined (or not) at a second beamsplitter BS$_2$ before detection. If the second beamsplitter is present we observe interference fringes, indicating the photon behaved as a wave, traveling both arms of the MZI. If BS$_2$ is absent, we randomly register, with probability $\frac 1 2$, a click in only one of the two detectors, concluding that the photon travelled along a single arm, showing particle properties.

This contradictory behaviour prompted Wheeler to formulate the delayed-choice experiment \cite{wheeler_dc, wheeler_wz, leggett, englert, exp_dc}. In Wheeler's delayed-choice experiment one randomly chooses whether or not to insert the second beamsplitter when the photon is already inside the interferometer and before it reaches BS$_2$ (Figure \ref{delayed_ch}a). The rationale behind the delayed choice is to avoid a possible causal link between the experimental setup and photon's behaviour: the photon should not ``know'' beforehand if it has to behave like a particle or like a wave. The choice of inserting or removing BS$_2$ is classically controlled by a random number generator.

In this article we examine what happens if we replace this classical control with a quantum device. This enables us to extend Wheeler's {\em gedanken} experiment to a quantum delayed-choice. Quantum elements in various experimental set-ups were proposed in the past \cite{penrose}. In order to understand the transition from a classical to a quantum controlling element it is insightful to reframe the delayed-choice experiment in terms of quantum networks \cite{barenco, nielsen}. A quantum network model enables us to analyze the {\em gedanken} experiment at a higher level of abstraction and to understand the information flow between different subsystems. The delayed-choice  experiment is equivalent to the quantum network in Figure \ref{delayed_ch}(b), where Hadamard gates $H$ play the role of beamsplitters; we call the top (black) line {\em the photon} and the bottom (red) line {\em the ancilla}. The quantum random number generator is modelled by an ancilla prepared in the equal-superposition state $\ket{+}= \frac{1}{\sqrt{2}}(\ket{0}+ \ket{1})$, then measured; the result of this measurement (0 or 1) controls if BS$_2$ is inserted or not. The classical control after the measurement of the ancilla in Figure \ref{delayed_ch}(b) is equivalent to a quantum control before the measurement of the ancilla, Figure \ref{delayed_ch}(c). This seemingly innocuous observation radically changes the setup and has two profound implications. First, since now we have a quantum beamsplitter in superposition of being present or absent, the interferometer is in a superposition of being closed or open. Following Wheeler's interpretation of the experiment \cite{wheeler_wz}, this forces the photon to be in a superposition of particle and wave at the same time.

\begin{figure}[floatfix]
\putfig{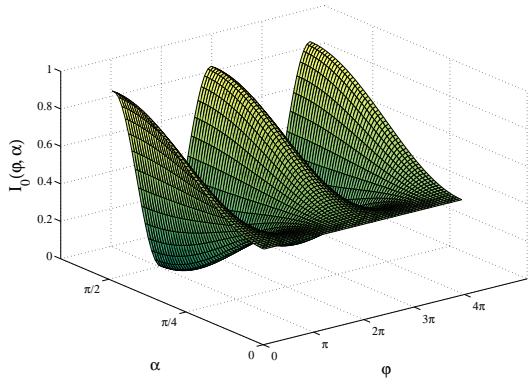}{8}
\caption{Morphing behaviour between particle ($\alpha=0$) and wave ($\alpha= \pi/2$).}
\label{morph}
\end{figure}

Second, and more important, a quantum control allows us to reverse the temporal order of the measurements. We can now detect the photon before the ancilla, i.e., before choosing if the interferometer is open or closed. This implies that we can choose if the photon behaves as a particle or as a wave {\em after} it has been already detected (post-selection). Consequently, this avoids the experimentally demanding requirement of an ultrafast switch necessary in the classical delayed-choice experiment \cite{exp_dc}. A quantum control thus allows us to explore a regime outside the classical realm: in any classically-controlled experiment the choice of inserting or not the second beamsplitter has to be made before the photon is detected. Since the photon and the ancilla interact at the $C(H)$ gate, the ancilla is always prepared before the photon reaches BS$_2$.

In Figure \ref{delayed_ch}(c), the photon--ancilla system starts in the state $\ket{00}$ and at the end of the network the final state is
\be
\ket{\psi}= \tfrac{1}{\sqrt{2}} (\ket{{\sf particle}}\ket{0}+ \ket{{\sf wave}}\ket{1})
\label{psi_f}
\ee
where the wavefunctions $\ket{{\sf particle}}= \tfrac{1}{\sqrt{2}}(\ket{0}+ e^{i\varphi}\ket{1})$ and $\ket{{\sf wave}}= e^{i\varphi/2}(\cos\tfrac{\varphi}{2}\ket{0}- i \sin\tfrac{\varphi}{2}\ket{1})$ describe particle and wave behaviour, respectively. The two states are in general not orthogonal $\braket{{\sf particle}}{{\sf wave}}= \tfrac{1}{\sqrt{2}} \cos\varphi$, except for $\varphi= \pm \pi/2$. Eq.~(\ref{psi_f}) implies that if the ancilla is measured to be $\ket{0}$ ($\ket{1}$), the interferometer is open (closed) and the photon behaves like a particle (wave). The interference pattern measured by the photon detector $D_0$ is $I_0(\varphi)= \tr(\rho_1 \ket{0}\bra{0})$, with $\rho_1= \tr_2\ket{\psi}\bra{\psi}= \frac 1 2 (\ket{{\sf particle}}\bra{{\sf particle}}+ \ket{{\sf wave}}\bra{{\sf wave}})$ the reduced density matrix of the photon. The visibility of the interference pattern is $\V= (I_{max}- I_{min})/(I_{max}+ I_{min})$, where the min/max values are calculated with respect to $\varphi$. If the interferometer is closed, the photon shows a wavelike behaviour with $I_w(\varphi)= \cos^2\frac \varphi 2$ and visibility $\V= 1$. For an open interferometer the photon behaves like a particle and $I_p(\varphi)=\frac 1 2$, resulting in $\V= 0$. For the entangled state (\ref{psi_f}) the result is
\be
I_0(\varphi)= \tfrac{1}{2} [ I_p(\varphi)+ I_w(\varphi) ]= \tfrac{1}{2} + \tfrac{1}{4}\cos\varphi
\ee
Without correlating the photon data with the ancilla we observe an interference pattern with reduced visibility $\V= \frac 1 2$: the photon has a mixed behaviour between a particle and a wave. On the other hand, if we do correlate the photon with the ancilla we observe either a perfect wave-like behaviour (ancilla $\ket{1}$) or a particle-like one (ancilla $\ket{0}$). Contrary to Bohr's opinion, we do not have to change the experimental setup in order to measure complementary properties -- we can measure both properties in a single experiment, provided that a component of the apparatus is a quantum object in a superposition state. The behaviour is post-selected by the experimenter after the photon has been detected, by correlating the data with the appropriate value of the ancilla \cite{kim}.

The photon in state $\ket{\psi}$ exhibits both wave and particle behaviour with equal probability. It is insightful to generalize this result to an arbitrary superposition. We achieve this by preparing the ancilla in the state $\cos\alpha \ket{0}+ \sin\alpha \ket{1}$ before interacting with the photon (Figure \ref{delayed_ch}d). In the classical setup (Figure \ref{delayed_ch}a) this choice corresponds to a biased random number generator which outputs $0$ with probability $\cos^2 \alpha$. The final state becomes
\be
\ket{\psi'}= \cos\alpha \ket{{\sf particle}}\ket{0}+ \sin\alpha \ket{{\sf wave}}\ket{1}
\label{psi_f2}
\ee
and the photon detector $D_0$ now measures:
\be
I_0(\varphi, \alpha)= I_p(\varphi) \cos^2\alpha + I_w(\varphi) \sin^2\alpha
\label{I0phi_alpha}
\ee
with the corresponding visibility $\V= \sin^2\alpha$. Thus, by varying $\alpha$ we have the ability to modify continuously the interference pattern -- we have a morphing behaviour between a particle at $\alpha=0$ and a wave at $\alpha= \pi/2$ (Figure \ref{morph}).

This continuously varying behaviour (morphing) raises questions about the classical picture of a photon as either a particle or a wave. A quantum beamsplitter transcends the ``particle-or-wave'' dichotomy and enables to prepare the photon in a superposition of both. For example, by measuring the ancilla controlling the beamsplitter in the $\ket{\pm}$ basis, the photon state becomes $\cos\alpha \ket{{\sf particle}} \pm \sin\alpha \ket{{\sf wave}}$, a superposition without a classical analog.

The introduction of a quantum control (i.e., quantum beamsplitter) allows us to answer an important question: {\em Can a hidden-variable (HV) theory, in which ``particle'' and ``wave'' are realistic properties, explain the delayed-choice experiment?} Such a model should satisfy two conditions: (i) it should reproduce the quantum mechanical (QM) statistics, and (ii) for a given photon the property of being a ``particle" or a ``wave'' is intrinsic, i.e., does not change during its lifetime. The second condition is very important, since it selects from the existing HV theories \cite{branyan} reproducing the QM statistics those models having meaningful notions of ``particle'' and ``wave''. Moreover, a quantum control potentially introduces new routes for causal influence, making the HV analysis \cite{branyan} more subtle. In the basis $a\otimes b= (00, 01, 10, 11)$ the statistics for the joint measurements of the photon $a$ and ancilla $b$ in the state (\ref{psi_f2}) is:
\be
p(a,b)= \big(\half \cos^2\! \alpha,\, \sin^2\! \alpha \cos^2\!\tfrac{\varphi}{2}, \half \cos^2\! \alpha,\, \sin^2\! \alpha \sin^2\!\tfrac{\varphi}{2} \big)
\label{pab_qm}
\ee
We show that there is no satisfactory HV model reproducing the statistics $p(a,b)$ and in which ``particle'' and ``waves'' are realistic properties. One can assume that the source randomly emits, with some probability, particle- or wave-like photons. However, in order to have the statistics $p(a,b)$ these ``photons'' show an inconsistent behaviour: in an open interferometer waves obey a particle statistics and in a closed interferometer particles behave like waves, showing interference. Consequently, the properties ``wave'' and ``particle'' become meaningless.

\noindent{\em Proof:} We assume the photon has an extra degree of freedom $\lambda$ (the hidden variable) corresponding to a particle-like ($\lambda=\p$) or a wave-like ($\lambda=\w$) behaviour. We also assume the standard conditions for probability distributions; for all variables $i,j$ we have: (i) $p(i)= \sum_j p(i,j)$ (marginals) and (ii) $p(i,j)= p(i|j) p(j)= p(j|i) p(i)$ (conditionals).

In this HV model the probability distribution $p(a,b)$ is the marginal of a distribution involving the hidden variable $\lambda$, namely $p(a,b)= \sum_\lambda p(a, b, \lambda)$, with $p(a,b,\lambda)$ unknown. We decompose this probability as $p(a, b, \lambda)= p(a| b, \lambda)\, p(b| \lambda)\, p(\lambda)$, by replacing the seven parameters $p(a, b, \lambda)$ with another seven functions (the probabilities in the rhs have four, two and respectively, one free parameter). This decomposition is appealing as the new functions are physically intuitive, unlike $p(a, b, \lambda)$. Thus we have
\be
p(a,b)= \sum_\lambda p(a| b, \lambda)\, p(b| \lambda)\, p(\lambda)
\label{p_ab}
\ee

Two of the conditional distributions $p(a|b,\lambda)$ are constrained by the expectation of how particles (waves) behave in open (closed) interferometers. Consistent with our previous definition, a particle in an open interferometer ($b=0$) has the statistics
\be
p(a|b=0, \lambda=\p)= \big(\half, \half \big)
\label{p_and_w.1}
\ee
whereas a wave in a closed MZI ($b=1$) shows interference:
\be
p(a|b=1,\lambda=\w)= \big( \cos^2\!\tfrac{\varphi}{2},  \sin^2\!\tfrac{\varphi}{2}\big)
\label{p_and_w.2}
\ee
The other two conditional probabilities specify the behaviour of a wave ($\lambda=\w$) in a open ($b=0$) interferometer and of a particle ($\lambda=\p$) in a closed ($b=1$) one. We denote these two unknown distributions by $x$ and $y$, respectively
\ba
\nonumber p(a|b=0, \lambda=\w)&=& (x, 1-x) \\
\nonumber p(a|b=1, \lambda=\p)&=& (y, 1-y)
\ea
The probability distribution of the ancilla $p(b)$ is obtained from eq.~(\ref{pab_qm}) as the marginal of $p(a,b)$
\be
p(b)= (\cos^2 \alpha, \sin^2 \alpha)
\ee
By freely choosing $\alpha$ at the preparation stage we modify $p(b)$, a fact which will prove crucial later.

For $\lambda$ we assume that the source randomly emits particle- or wave-like photons with probability $f$ and $1-f$, respectively:
\[
p(\lambda)= (f, 1-f)
\]
The remaining two variables are the conditional probability distributions of the ancilla $b$ and the hidden variable $\lambda$:
\ba
\nonumber
p(b| \lambda=\p)&=& (z, 1-z) \\
\nonumber
p(b| \lambda=\w)&=& (v, 1-v)
\ea
satisfying the consistency condition $p(b)= \sum_\lambda p(b| \lambda) p(\lambda)$. From eqs.~(\ref{pab_qm}), (\ref{p_ab}) with the constraints (\ref{p_and_w.1}), (\ref{p_and_w.2}) we obtain:
\begin{eqnarray}
v(1-f)(x- \half)&=& 0 \label{3.1} \\
f(1-z)(y- \cos^2 \tfrac{\varphi}{2})&=& 0 \label{3.2}\\
zf+ v(1-f)- \cos^2\alpha &=& 0 \label{3.3}
\end{eqnarray}
As $\alpha$ is arbitrary, we disregard the cases $v=0, f=0$, implying $\cos^2 \alpha=0$ and $f=1, z=1$, giving $\cos^2 \alpha=1$.

Five of the remaining non-trivial solutions have either $x= \half$ or $y= \cos^2 \tfrac{\varphi}{2}$ (or both). The solution $x= \half$ means that waves in open interferometers have a particle statistics, $p(a| b=0, \lambda=\w)= (\half, \half)$. The second solution $y= \cos^2 \tfrac{\varphi}{2}$ implies that particles in closed interferometers behave like waves, $p(a| b=1, \lambda=\p)= (\cos^2 \tfrac{\varphi}{2}, \sin^2 \tfrac{\varphi}{2})$. None of these solutions is acceptable, as particles and waves show an inconsistent behaviour: waves in open interferometers have particle statistics and particles in closed interferometers show interference. The last solution is:
\be
v=0,\ \  z=1,\ \  f=\cos^2 \alpha
\ee
with $x,y$ undetermined. In other words, the source randomly emits particles and waves with a distribution $p(\lambda)= (\cos^2 \alpha, \sin^2 \alpha)$ identical to the probability distribution $p(b)$ of the ancilla. Moreover, whenever the source emits a particle-like photon the ancilla is found to be 0, $p(b| \lambda=\p)= (1,0)$ and the interferometer is open. On the other hand, when it emits a wave-like photon the ancilla is measured as 1, $p(b| \lambda=\w)= (0,1)$, so the interferometer is closed. The hidden-variable $\lambda$ and the ancilla $b$ are perfectly correlated, $p(b| \lambda)= \delta_{\lambda \p}\delta_{b 0}+ \delta_{\lambda \w}\delta_{b 1}$.

The paradox is now revealed: although the hidden-variable completely determines the value of the ancilla, the probability distribution $p(\lambda)$ is identical to $p(b)$ which is set by the experimenter preparing $\alpha$. To explain this, we need to enlarge the HV theory in order to include also the setting $\alpha$, resulting in a second-order HV theory (deemed unacceptable by Bell \cite{bell-book}). This invites an induction {\em ad infinitum} procedure, in which we introduce a second (and third etc) ancilla in order to offset the causality between the source and the preparation of the lower-order ancilla. In this scenario we have a delayed-delayed-\ldots-choice experiment all the way down. Occam's razor compels us to cut this infinite chain to the first link. In conclusion, if the hidden-variable $\lambda$ completely determines $b$, then $\lambda$ itself cannot be determined by the setting $\alpha$ preparing $b$.

To summarize, we have shown that any HV theory that reproduces the QM statistics $p(a,b)$ and agrees with natural definitions of particle and wave behaviour, either assumes wave-particle duality (which was supposed to abolish in the first place) or introduces higher-order HV theories. $\hfill \Box$

The definition of ``particle" (``wave'') used above is based on the observed statistics in a open (closed) interferometer (\ref{p_and_w.1})--(\ref{p_and_w.2}), as this is the only meaningful possibility in a probabilistic theory as QM. As noted before, from a classical perspective there is still an ontological tension between the observed interference and the detection of individual photons, one by one, by clicks in the detectors.

In conclusion, we proposed and analysed a quantum version of Wheeler's delayed-choice experiment. This has several important consequences. First, the photon shows a morphing behaviour between ``particle" and ``wave". This further supports the conclusion that ``particle" and ``wave" are not realistic properties but merely reflect how we 'look' at the photon; such behaviour is a direct consequence of a quantum beam-splitter and cannot be revealed in a classical setup. Second, the classical choice {\em particle vs.~wave} can be made after the photon has been already detected, by correlating the photon data with the measured value of the ancilla (post-selection). We have shown that complementary phenomena can be observed with a single experimental setup, provided that a component of the apparatus is a quantum device in a superposition state. Our result suggests a reinterpretation of the complementarity principle -- instead of complementarity of experimental setups (Bohr's view) we have complementarity of experimental data. We anticipate quantum controls will play an important role in re-assessing other experiments in foundations of quantum mechanics, particularly Bell-inequality tests \cite{bell65, aspect82}.

Discussing the delayed-choice experiment, Wheeler concludes: ``In this sense, we have a strange inversion of the normal order of time. We, now, by moving the mirror in or out have an unavoidable effect on what we have a right to say about the {\em already} past history of that photon'' \cite{wheeler_wz}. We disagree with this interpretation. There is no inversion of the normal order of time -- in our case we measure the photon {\em before} the ancilla deciding the experimental setup (open or closed interferometer). It is only after we interpret the photon data, by correlating them with the results of the ancilla, that either a particle- or wave-like behaviour emerges: {\em behaviour is in the eye of the observer}.

\noindent {\bf Acknowledgements} We thank R.~Colbeck, A.~Leggett, G.~Milburn, A.~Peruzzo, H.~Price, S.~Rebi\'c, T.~Rudolph and J.~Twamley for comments and discussions. This work was supported by the ARC Centre for Quantum Computer Technology and EC Project QUANTIP 244026.

\end{document}